\documentclass[twocolumn]{aastex7}
\usepackage{hyperref}
\usepackage{CJK}
\usepackage{graphicx}	
\usepackage{amsmath}	
\usepackage{amssymb}	
\usepackage{float}
\usepackage{graphicx}
\usepackage{subfigure}
\usepackage{multirow}

\usepackage{xcolor} 

\usepackage{hyperref}  

\submitjournal{xxx}
\begin{document}
\begin{CJK*}{UTF8}{gbsn}

\title{Antarctic Infrared Binocular Telescope: Early Data Release of observations in the 1.4 \textmu m water-vapor-absorption band}
\shortauthors{Lin et al.}

\correspondingauthor{Bin Ma}
\email{mabin3@mail.sysu.edu.cn}

\author[0009-0007-2850-9908]{Pu Lin({\CJKfamily{gbsn}林浦})}
\affiliation{School of Physics and Astronomy, Sun Yat-sen University, Zhuhai 519082, People's Republic of China}
\email{linp@mail2.sysu.edu.cn}

\author[0009-0007-2408-9221]{Haonan Yang({\CJKfamily{gbsn}杨浩楠})}
\affiliation{School of Physics and Astronomy, Sun Yat-sen University, Zhuhai 519082, People's Republic of China}
\email{yanghn8@mail2.sysu.edu.cn}

\author[0000-0002-6077-6287]{Bin Ma({\CJKfamily{gbsn}马斌})} 
\affiliation{School of Physics and Astronomy, Sun Yat-sen University, Zhuhai 519082, People's Republic of China}
\affiliation{CSST Science Center for the Guangdong-Hong Kong-Macau Greater Bay Area, Zhuhai 519082, People's Republic of China}
\email{mabin3@mail.sysu.edu.cn}

\author[0009-0000-6955-0594]{Jinji Li({\CJKfamily{gbsn}李晋基})}
\affiliation{School of Physics and Astronomy, Sun Yat-sen University, Zhuhai 519082, People's Republic of China}
\email{lijj328@mail2.sysu.edu.cn}

\author[0009-0006-9178-4024]{Haoran Zhang({\CJKfamily{gbsn}张皓然})}
\affiliation{School of Physics and Astronomy, Sun Yat-sen University, Zhuhai 519082, People's Republic of China}
\email{zhanghr33@mail2.sysu.edu.cn}

\author[0000-0003-1412-2028]{Michael C.~B. Ashley}
\affiliation{School of Physics, University of New South Wales, Sydney, NSW 2052, Australia}
\email{m.ashley@unsw.edu.au}

\author[0009-0003-5592-3734]{Zhongnan Dong({\CJKfamily{gbsn}董仲南})}
\affiliation{School of Physics and Astronomy, Sun Yat-sen University, Zhuhai 519082, People's Republic of China}
\affiliation{National Astronomical Observatories, Chinese Academy of Sciences, Beiing 100101, China}
\email{dongzhn@mail2.sysu.edu.cn}

\author[]{Lu Feng({\CJKfamily{gbsn}冯麓})}
\affiliation{National Astronomical Observatories, Chinese Academy of Sciences, Beiing 100101, China}
\email{jacobfeng@bao.ac.cn}

\author[0009-0003-0689-0067]{Wei Huang({\CJKfamily{gbsn}黄伟})}
\affiliation{School of Physics and Astronomy, Sun Yat-sen University, Zhuhai 519082, People's Republic of China}
\email{huangw327@mail2.sysu.edu.cn}

\author[0000-0003-3317-4771]{Yi Hu({\CJKfamily{gbsn}胡义})}
\affiliation{National Astronomical Observatories, Chinese Academy of Sciences, Beiing 100101, China}
\email{huyi.naoc@gmail.com}

\author[0000-0002-6796-124X]{Zhaohui Shang({\CJKfamily{gbsn}商朝晖})}
\affiliation{National Astronomical Observatories, Chinese Academy of Sciences, Beiing 100101, China}
\email{zshang@gmail.com}

\author[0009-0003-0141-5793]{Yun Shi({\CJKfamily{gbsn}史韵})}
\affiliation{School of Physics and Astronomy, Sun Yat-sen University, Zhuhai 519082, People's Republic of China}
\email{shiy228@mail2.sysu.edu.cn}

\author[0000-0003-2775-3523]{Shijie Sun({\CJKfamily{gbsn}孙士杰})}
\affiliation{National Astronomical Observatories, Chinese Academy of Sciences, Beiing 100101, China}
\email{sshj@nao.cas.cn}

\author[0000-0003-4147-8759]{Xu Yang({\CJKfamily{gbsn}杨栩})}
\affiliation{National Astronomical Observatories, Chinese Academy of Sciences, Beiing 100101, China}
\email{xuyang@nao.cas.cn}

\author[0000-0002-1086-7922]{Yong Zhang({\CJKfamily{gbsn}张泳})}
\affiliation{School of Physics and Astronomy, Sun Yat-sen University, Zhuhai 519082, People's Republic of China}
\affiliation{CSST Science Center for the Guangdong-Hong Kong-Macau Greater Bay Area, Zhuhai 519082, People's Republic of China}
\email{zhangyong5@mail.sysu.edu.cn}

\begin{abstract}

Ground-based observations around 1.4 \textmu m are normally limited by strong absorption of telluric water-vapor.
However, Dome A, Antarctica has exceptionally dry conditions that offer a unique opportunity for observations in this band.
We designed a new filter covering 1.34--1.48 \textmu m, namely $W'$, and installed it on the Antarctic Infrared Binocular Telescope (AIRBT) at Dome A in 2025.
AIRBT comprises two identical 15 cm optical tube assemblies and two InGaAs cameras equipped with $J$ and $W'$ filters, respectively. 
With this Early Data Release (EDR), we aim to evaluate the performance of the $W'$ band at Dome A to observe objects with water-vapor features.
This EDR covers $\thicksim 20 \ \mathrm{deg^2}$ in the Galactic plane using $\thicksim 20,000$ images in three nights.
For 2 s exposures, the 5 $\sigma$ limiting magnitude histogram peaks at $J \thicksim 11.5$ mag (Vega) and $W' \thicksim 9.9$ mag, respectively.
The $J-W'$ vs $J-H$ color-color diagram distinguishes ultracool candidates with water-vapor-absorption features from reddened early type stars. 
Furthermore, later-type stars tend to exhibit stronger water-vapor absorption.
Some sources show larger $\Delta W'$ than $\Delta J$ across the three nights, which we attribute to variations of their water-vapor-absorption depth.
We conclude that it will be efficient to search for ultracool stars and estimate their spectral subtypes using $W'$ band imaging at Dome A,
where the atmospheric transmission is high and stable.

\end{abstract}

\keywords{Catalogs (205) --- Near infrared astronomy (1093) --- Stellar atmospheres (1584) --- Photometry (1234)}

\section{Introduction}

Water plays a crucial role in the study of astronomy.
Observations show that
water is widely found in the universe
\citep[e.g.,][]{2013ChRv..113.9043V}.
In cool giants, water-vapor-absorption features reflect the structure and molecular composition of their outer atmospheres.
Spectroscopic studies of Mira variables have shown that water-vapor in their atmospheres influences atmospheric convection and pulsational behavior
\citep{wittkowski2016near}.
Water is also present in molecular clouds that undergo star formation and plays an important role in their chemistry and evolution
\citep{1997ApJ...486..316B, 2007ARA&A..45..339B, 2014prpl.conf..835V, van2021water}.
In exoplanet atmospheres, water influences dynamics and solid accretion during planetary evolution
\citep{2006Icar..181..178C,2013A&A...552A.137R};
moreover, water's abundance and chemical form strongly affect a planet's habitability and are crucial to studies of the origin of life
\citep{2023ASPC..534.1031K}.

Water exists as vapor in the atmosphere of ultracool stars, brown dwarfs, and exoplanets.
Consequently, water produces many wide and deep absorption features in infrared spectra.
Spectroscopic observations show that cool giants and dwarfs exhibit absorption features near 1.4 \textmu m caused by water-vapor,
and that the lower the temperature (i.e. the later the spectral type), the deeper the absorption\citep{2009ApJS..185..289R}.
Therefore, these features can be used for batch classification, candidate search, and spectral type estimation.
In exoplanet atmospheric studies, water-vapor-absorption is a crucial tool to find potential habitable planets.
For example, a JWST transit spectroscopic observation of WASP-96b revealed water-vapor-absorption peaks in the infrared,
confirming the presence of water-vapor in its atmosphere
\citep{10.1093/mnras/stad1547, 10.1093/mnras/stad1762}.
Through water-vapor-absorption near 1.4 \textmu m, HST transit observations detected water-vapor in the atmosphere of the hot Jupiter HD 189733b \citep{fraine2014water} and the Neptune-sized exoplanet HAT-P-11b \citep{mccullough2014water}.

Observing water-vapor features in astrophysical objects from the surface of the Earth is challenging due to the water-vapor content in the Earth's atmosphere and particularly by its variability. That is why most current observations in the promising 1.4 \textmu m band rely mainly on space telescopes
such as HST/WFC3
with F139M/F140W
\citep{dressel2016wide}
filters and JWST/NIRCam with F140M/F150W
\footnote{\url{https://jwst-docs.stsci.edu/jwst-near-infrared-camera/nircam-instrumentation/nircam-filters}}.
\cite{allers2020novel} have developed a $W$ filter with a central wavelength of 1.45 \textmu m
and a bandpass width of 6\% ($\thicksim$ 87 nm), which covers part of the water-vapor-absorption region, while avoiding
the area with the strongest atmospheric absorption at Mauna Kea.
With the $W$ imaging, they efficiently distinguished late M and L dwarfs from reddened background stars in star-forming regions.
They estimated spectral types with a precision of 1.4 subtypes on the basis of the absorption depth.

\begin{figure*}
    \centering
    \includegraphics[width=\linewidth]{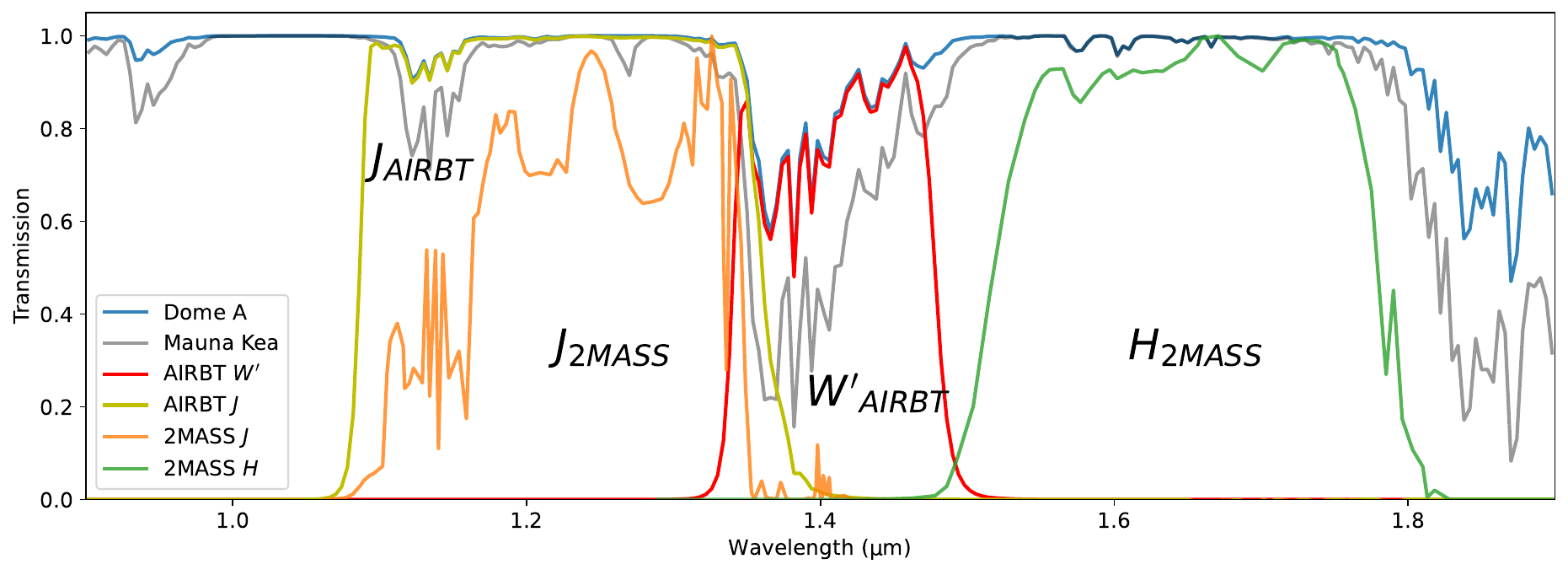}
    \caption{Theoretical atmosphere transmission at Dome A (blue, median PWV = 0.141 mm, airmass = 1.0)
    and Mauna Kea (gray, median PWV = 1.6 mm, airmass = 1.0) at a 4 nm resolution,
    and filter transmission curves of AIRBT $J$, $W'$ filters and 2MASS $J$, $H$ filters.
    All filter transmission curves include the effect of atmosphere transmission.}
    \label{fig:filter}
\end{figure*}

Dome A in Antarctica is an ideal site for ground-based observations
\citep{shang2020astronomy, ma2020night, yang2021cloud, hu2018meteorological},
and especially its extremely dry condition, an almost perfect site for 1.4 \textmu m observation
because of its extremely dry air.
The median precipitable water vapor (PWV) is about 0.13--0.14 mm \citep{yang2010exceptional, Sims_2012, kuo2017assessments, shi2016terahertz},
yielding a high atmospheric transmission near 1.4 \textmu m \citep{Sims_2012}.
To verify the performance of this new window at Dome A, we designed a new filter, $W'$, covering 1.34--1.48 \textmu m
\citep{zhang2024customizing}.
Compared with the original $W$ filter at Mauna Kea \citep{allers2020novel}, our
$W'$ filter is broader and more transparent at Dome A, which will improve observational efficiency.
Figure \ref{fig:filter} compares the atmosphere transmission at Dome A \citep{zhang2024customizing} and Mauna
Kea\footnote{\url{https://www.gemini.edu/observing/telescopes-and-sites/sites\%23MKWV\#Transmission}}
together with the filter transmission curve of AIRBT and 2MASS \citep{cohen2003spectral}.

In 2025 we installed the $W'$ filter in one of the telescopes in the Antarctic Infrared Binocular Telescope (AIRBT) \citep{dong2025antarctic}, replacing the original $H'$ filter.
Our observations with the $W'$ filter aim to search for and study ultracool giants and dwarfs by detecting their
water-vapor-absorption features, and to estimate their spectral types.
In this paper, we present an Early Data Release (EDR) of AIRBT observations in 2025,
and explore the scientific potential of the $W'$ filter.
The paper is organized as follows:
Sect. \ref{sec:instrument} introduces AIRBT and the observation data in the EDR;
Sect. \ref{sec:data} describes the data reduction and photometric calibration;
Sect. \ref{sec:results} presents photometric results, data products, and direct analysis from catalogs,
including using the color-color diagram and variability to search for candidates with water-vapor-absorption;
Sect. \ref{sec:discussion} discusses the correlation between water-vapor-absorption and spectral type,
and examines the atmospheric transmission stability at Dome A;
Sect. \ref{sec:summary} provides a summary and our future plans.


\section{Telescope and observations}
\label{sec:instrument}

\subsection{Telescope overview}

\begin{figure*}
    \centering
    \includegraphics[width=\linewidth]{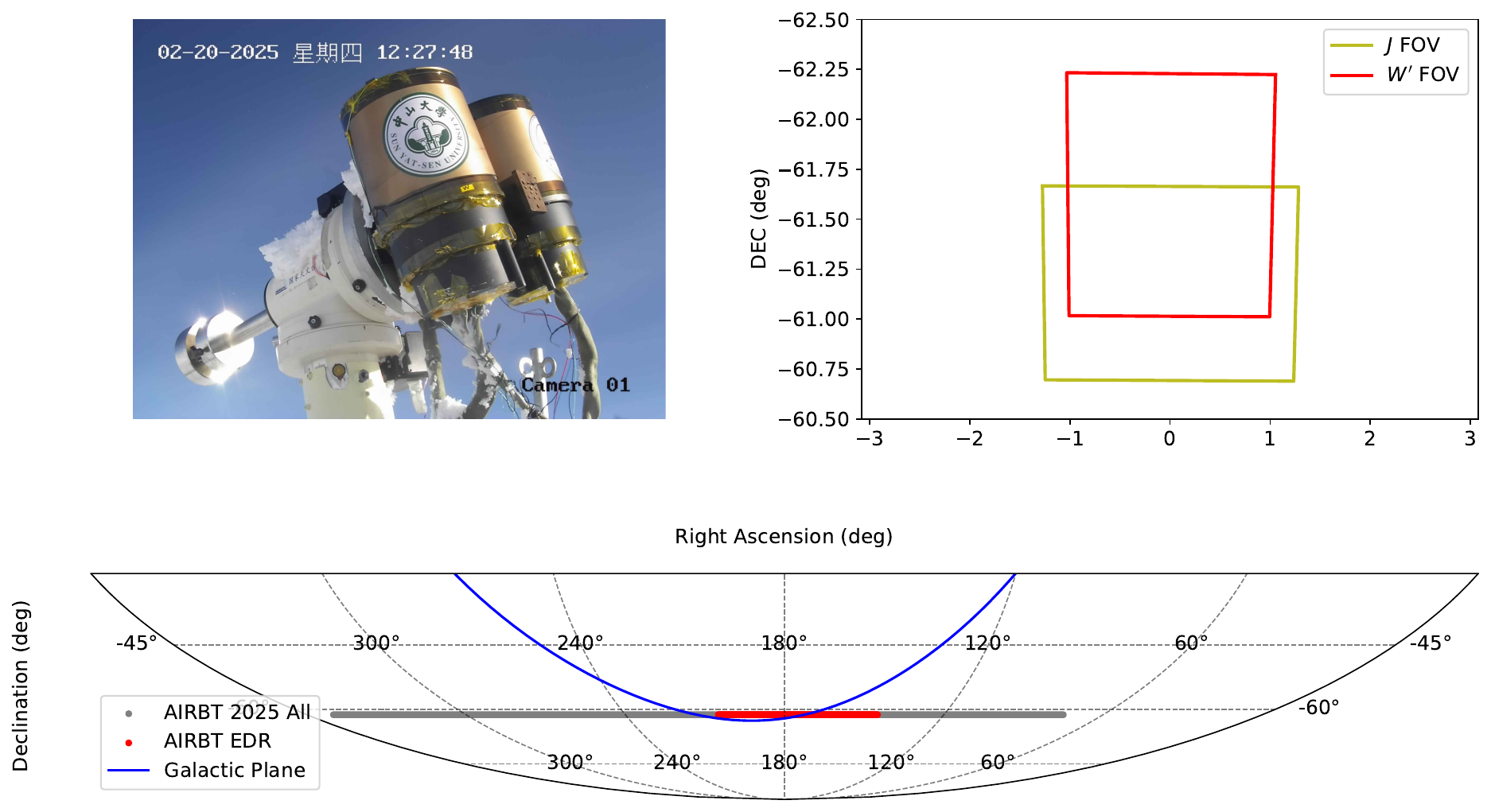}
    \caption{Upper Left: The Antarctica Infrared Binocular Telescope (AIRBT) at Dome A, Antarctica.
              Upper Right: The FoVs of the $J$ (shown in yellow) and $W'$ (shown in red) bands.
              The overlapping area is about $\mathrm{512 \times 360 \ pixel}$
              and covers $\delta=-61.7^\circ \thicksim -61.0^\circ$.
              Lower: The sky coverage of the AIRBT Early Data Release (red) and the full data set  (gray) in 2025.
              The blue line is the Galactic plane.
              }
    \label{fig:AIRBT}
\end{figure*}

AIRBT (see upper left Fig. \ref{fig:AIRBT}) comprises two 15 cm optical tube assemblies in a Ritchey-Chrétien (RC) design with a lens corrector and a f/3 focal ratio.
Each focal plane array consists of a commercial InGaAs camera with $\mathrm{640 \times 512 \ pixel}$ 15 \textmu m pixels,
giving a pixel scale of $\mathrm{6.84\arcsec \ pixel^{-1}}$ and a field-of-view (FoV) of $1.22^\circ \times 0.97^\circ$.

After technical validation in China in October 2022 \citep{dong2025antarctic},
AIRBT was installed at Dome A, Antarctica in January 2023 by
the 39th China National Antarctic Research Expedition (CHINARE 39).
After maintenance in January 2024 by CHINARE 40, AIRBT carried out time domain surveys in $J$ and $H$ simultaneously.
During maintenance in January 2025 by CHINARE 41, AIRBT's $H$ filter was replaced with a $W'$ filter and began observations in the $J$ and $W'$ bands. All power and data transfer were provided through the PLATeau Observatory for Dome A
\citep[PLATO-A,][]{lawrence2009plato, 2010EAS....40...79A}
platform.

As the transmission curves show in Fig. \ref{fig:filter},
the $W'$ filter is specially designed to cover the entire water-vapor-absorption region near 1.4 \textmu m.
Additionally, compared with 2MASS, our $J$ filter is broader to increase SNR and its blue-ward side has larger efficiency because of atmosphere transmission at Dome A.

\subsection{Observations}

AIRBT's observations in the $J$ and $W'$ bands began in January 2025 
as tests of system stability and performance.
During February, with the Sun continuously above the horizon at Dome A, we were restricted to monitoring several bright stars.
From 16 March, with the night emerging, the telescope conducted continuous observations until 23 April, when a communication issue suspended operations.
During the night-time observations, the telescope was operated in drift scan mode due to mechanical issues with the mount,
so observations were restricted to the sky drifting naturally across the FoV.

We selected the Galactic plane as the primary area of interest due to high star density and strong interstellar extinction, which can highlight our advantages.
Strong interstellar extinction reddens stars to varying degrees,thereby reducing the reliability of spectral type identification using broadband colors.
In contrast, water-vapor-absorption features near 1.4 \textmu m are essentially unaffected by extinction,
thus offering a more reliable method for estimating spectral types.
To observe the Galactic plane
we fixed the pointing at declination of $\delta = -61.3^\circ$, while minimizing air mass by pointing almost due north; the image center was at $ALT = 358.265^\circ$ and $AZ = 71.375^\circ$,

The images were automatically captured, with
exposure times adjusted according to the background from daytime to nighttime.
The maximum exposure time was set to 2 s to balance PSF elongation and observation efficiency.
During the nighttime, a group of 200 exposures was captured at a time, with about 40~s overhead between groups.
During observations, we power on the Indium-Tin-Oxide (ITO) films coated on the windows to warm the windows and consequently prevent frost accumulation.
The ITO performance was remarkably effective (see Fig. \ref{fig:DomeA-air}) while having a negligible effect on image quality (see Fig. \ref{fig:hist}).

The $J$ and $W'$ detectors had different X-Y orientations because of mounting restrictions,
thus the FoVs of the $J$ and $W'$ bands did not fully overlap, as the upper right panel of Fig. \ref{fig:AIRBT} shows.
The overlapping area of $J$ and $W'$ bands covers $\delta=-61.7^\circ \thicksim -61.0^\circ$,
and a star drifting through it stays within it for about eight minutes.
The coverage of RA was dependent on the length of the night,
with $\alpha=145^\circ \thicksim 205^\circ$ for the first night (16 March 2025), and
$\alpha=75^\circ \thicksim 350^\circ$ for the last night (22 April 2025), as the gray markers show in the lower panel of Fig. \ref{fig:AIRBT}.

All raw data were stored at Dome A and partially downloaded via internet from the PLATO-A platform. The full data are scheduled to be retrieved by the next traverse.
In this paper, we selected three nights (16 March, 3 April, and 20 April 2025) for this Early Data Release (EDR); these nights correspond to 
the beginning, middle, and end of the observing period.
Our aims with the EDR are to evaluate the telescope's performance and to explore the scientific potential of the $W'$ filter.
The overlapping area of those three nights covers $\alpha=145^\circ \thicksim 205^\circ$ and lies entirely within the Galactic plane,
including $\thicksim$ 20,000 images and sky coverage of $\thicksim$ 20 deg$^2$,
as the red markers shown in the lower panel of Fig. \ref{fig:AIRBT}.


\section{Data reduction}
\label{sec:data}

In the image reduction, we subtracted the background,
divided by the twilight flat and the photometric flat (also known as the star flat),
and performed aperture photometry and astrometry.
To make reference stars for the $W'$ band, we transformed $J$ and $H$ magnitudes to $W'$ magnitudes as a preliminary photometric calibration.

\subsection{Preprocessing and aperture photometry}

\begin{figure*}
    \centering
    \includegraphics[width=\linewidth]{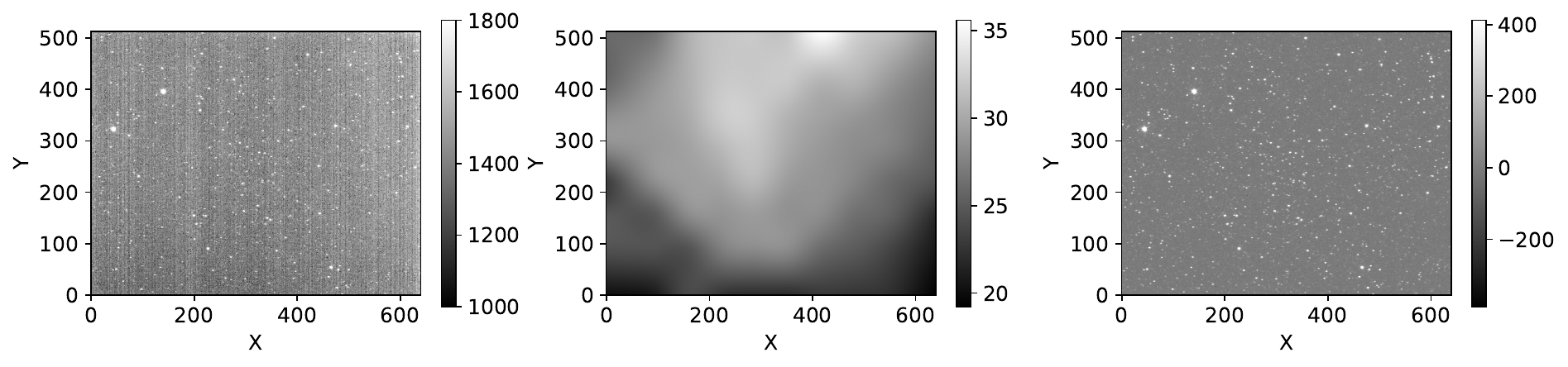}
    \caption{An example of raw  (left), sampled sky background residuals from SEP after first background subtraction (middle), and background-subtracted (right) image.}
    \label{fig:bkg_sub}
\end{figure*}

As shown in the left panel of Fig. \ref{fig:bkg_sub},
the background of the raw images exhibits significant noise,
including bias fixed patterns, dark current, warm pixels, and sky background nonuniformity.
The usual procedure to correct such background is to acquire bias and dark frames and subtract them from the raw images.
However, since the telescope was located at an unmanned location in the center of Antarctica and lacked a mechanical shutter,
we could not obtain dark and bias images.

To determine the background, we took the median of the raw images for each night after 3 $\sigma$ clipping to eliminate stars that naturally cross the field.
This background includes bias, dark current, and the sky background.
We subtracted the background image from the raw images taken during that night, and used the Python package SEP
\citep{Barbary2016}
to further sample and subtract the sky background residuals
in the right panel of Fig. \ref{fig:bkg_sub}.

\begin{figure*}
    \centering
    \includegraphics[width=\linewidth]{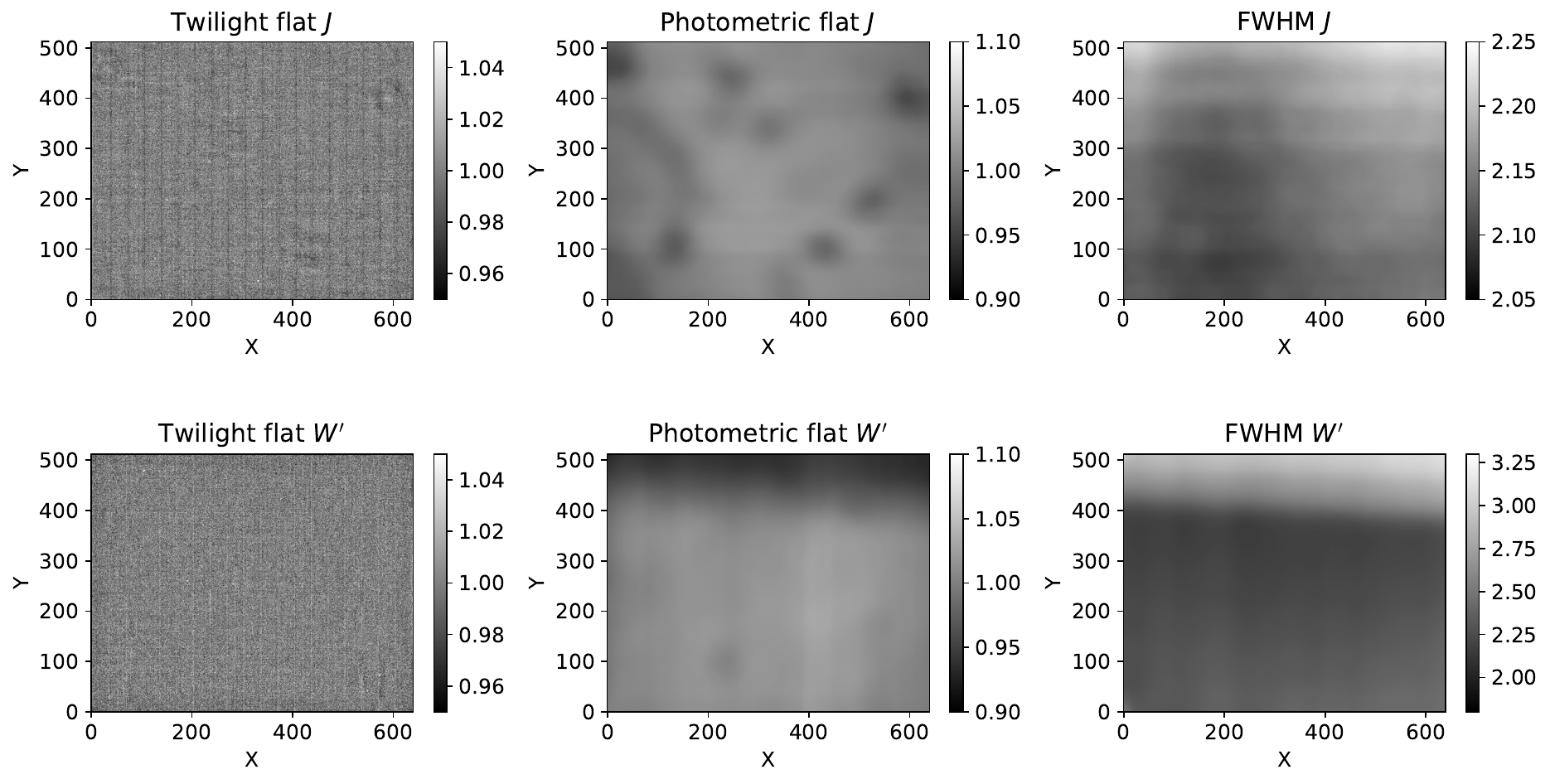}
    \caption{The upper (lower) panels show the twilight flat, photometric flat, and distribution map of the FWHM, respectively, for $J$ ($W'$) band.
            The dark spots in the $J$ band photometric flat are likely caused by ice crystals,
            while the striping in the $W'$ band photometric flat is correlated with the striping in the map of FWHM and may be introduced by the $W'$ filter,
            which will be inspected on the next expedition.}
    \label{fig:flat}
\end{figure*}

Both the twilight flat and the photometric flat were used in succession to correct spatial response nonuniformity of images.
To correct pixel-scale nonuniformity, we used the twilight flat, which was derived from the median of images taken at dawn and dusk.
We divided out the large-scale nonuniformity in the twilight flat using SEP
to reduce the effects from the twilight gradient and stray light.
Then, to correct large-scale nonuniformity, we used a photometric flat \citep[e.g.,][]{selman2004photometric}, which was computed by tracking the flux variations of stars as they drifted across the image. 
The photometric flat contained the information of image quality inhomogeneity and possible ice crystals on the lens.
Fig. \ref{fig:flat} shows the twilight flat, photometric flat, and the FWHM across the FoV.

We used SExtractor
\citep{1996A&AS..117..393B}
for source extraction and aperture photometry.
As shown in the right panels of Fig. \ref{fig:flat}, the FWHM was close to 2 pixels in the image.
In aperture photometry,
we adopted a 3 pixels aperture for faint stars and a 6 pixels aperture for bright stars,
to achieve the best signal-to-noise ratio by balancing the star flux and background contamination.

We used SCAMP
\citep{2006ASPC..351..112B}
to determine the World Coordinate System (WCS) parameters and
took the 2MASS catalog \citep{skrutskie2006two} as the astrometric reference.
Compared with 2MASS, our astrometric accuracy was 3.8\arcsec.

\subsection{Photometric calibration}
\label{sec:calibration}

As a newly designed filter, photometric calibration of $W'$ band is essential.
The usual procedure is to observe standard stars and use their fluxes for calibration.
However, as there are few standard stars within our current area,
in the current processing we derived preliminary $W'$ magnitudes by transforming the $J$ and $H$ magnitudes.
A refined calibration either based on standard stars or spectra from SPHEREx \citep{crill2025spherex} will be performed in future work.

\begin{figure*}
   \centering
   \includegraphics[width=\linewidth]{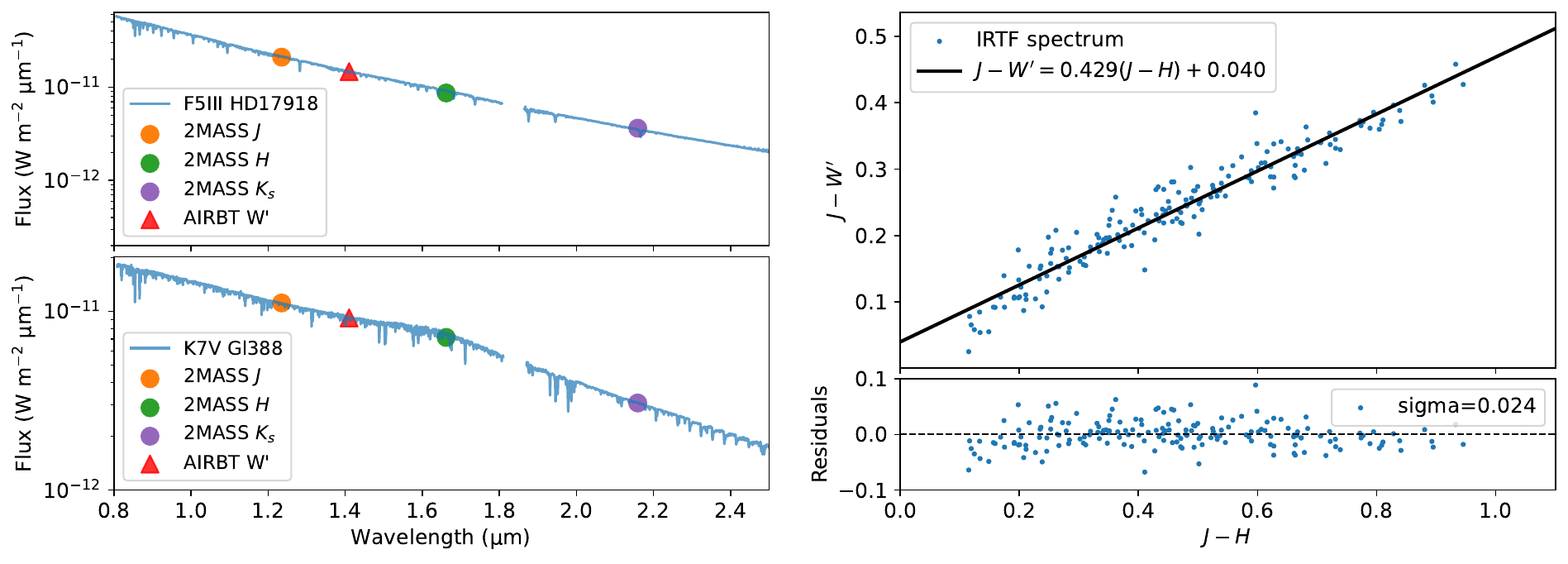}
   \caption{Transformation for $W'$ band. 
            The left upper (lower) panel is the spectrum of F5III HD17918 (K7V Gl388),
            the orange, green and purple points are fluxes converted from 2MASS $J$, $H$ and $K_s$;
            the red triangle is the AIRBT $W'$ flux.
            The right panel is $J-W'$ vs $J-H$ color-color diagram for stars earlier than K7 from the IRTF spectral library,
            and the black line is a linear fit of it, as the transformation for $W'$ band.}
   \label{fig:cal}
\end{figure*}

In the left panel of Fig. \ref{fig:cal}, the example spectra of stars earlier than K7
exhibit relatively featureless and straight (on a logarithmic flux scale) near-infrared continua,
particularly between the $J$ and $H$ bands.
This makes them suitable for a linear transformation from $J$ and $H$ magnitudes to $W'$ magnitudes.
We selected spectra of 186 stars earlier than K7 from the IRTF spectral library
\citep{2009ApJS..185..289R,2005ApJ...623.1115C,2003PASP..115..362R}
as reference stars.
For each spectrum, we computed the $W'$ magnitude by convolving the spectrum with the $W'$ filter transmission curve
and adopting the Vega flux as the photometric zeropoint.
We plot $J-W'$ vs $J-H$ color-color diagram in the right panel of Fig. \ref{fig:cal},
and derived a linear fitting as the transformation equation for $W'$ band:
\begin{equation}
    W'_{ref} = J - 0.429 \times (J-H) - 0.040.
    \label{eq:W_fit}
\end{equation}
The fitting has an RMS scatter of approximately 0.024 mag, which is sufficient for our preliminary analysis.
This transformation is independent of extinction, but only depends on the central wavelengths of the three bands.
With the central wavelength of 2MASS $J$, $H$ and AIRBT $W'$ filters, we derived a slope of
$(\lambda_J - \lambda_{W'})/(\lambda_J - \lambda_{H}) = 0.42$,
which agrees with our fitting in Eq. \ref{eq:W_fit} within the uncertainties.

For calibration of both the $J$ and $W'$ bands, we used the 2MASS catalog as a reference.
Since our pixel scale ($\mathrm{6.84 \arcsec \ pixel^{-1}}$) is much larger than
that of 2MASS ($\mathrm{1 \arcsec \ pixel^{-1}}$),
directly cross-matching would introduce significant errors.
For AIRBT bright sources with $J = 7 \thicksim 9$,
we find that about 10\% (40\%) of them are contaminated by more than one additional 2MASS source brighter than 10 (12) mag within their FWHM radius (2 pixels / 13.6\arcsec).
Therefore, for each AIRBT source, we searched all 2MASS sources within a radius, summed their fluxes,
and then converted it to magnitude as the reference.
Since our primary targets are bright stars with a 6 pixels aperture for photometry,
we set a matching radius to 3 pixels ($3 \times 6.84\arcsec$).
As this work represents a preliminary analysis, we adopted this simplified processing strategy.
A more detailed calibration procedure will be applied in future analyses of the full dataset,
including down-sampling the 2MASS images and re-performing photometry as a reference catalog.

\section{Results}
\label{sec:results}

In this section, we present the photometric performance of the $J$ and $W'$ bands images, and describe the EDR. With these data products, we achieve some initial results, including a color-color diagram and variability analysis, which enables a search for candidates with water-vapor-absorption features. 

\subsection{Data Products}

\begin{figure*}
    \centering
    \includegraphics[width=\linewidth]{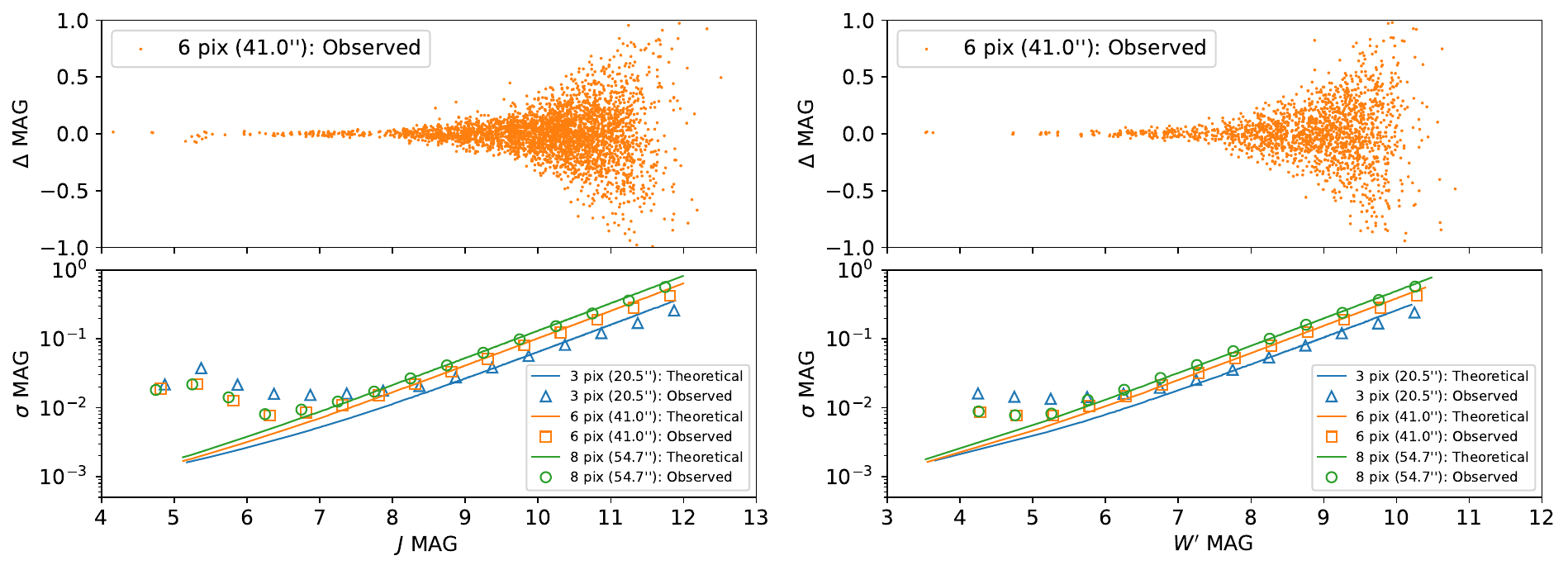}
    \caption{Photometric precision of the $J$(left) and $W'$(right) bands.
            For each band, the upper panel shows $\Delta$MAG vs MAG between pairs of two consecutive images,
            while the lower panel shows the photometric errors. The observed errors for three aperture sizes were calculated
            from the RMS of $\Delta$MAG in each 0.5 mag interval divided by $\sqrt{2}$, while the theoretical errors were from SExtractor.  
            In a 2 s exposure, the best photometric precision with a 6 pixels aperture reaches 8 mmag;
            with a 3 pixels aperture, the limiting magnitudes in the $J$ and $W'$ bands are 11.5 and 9.9, respectively.}
    \label{fig:magerr}
\end{figure*}

We evaluated our photometric precision by comparing pairs of two consecutive images (each of which is formed from 200 exposures, as previously discussed), which have almost identical conditions and hence the magnitude differences are dominated by photometric errors.
For each pair, we cross-matched sources with a separation of less than 1 pixel (6.84\arcsec)
and calculated the magnitude differences. 
Then in each 0.5 mag bin, we derived the magnitude differences RMS divided by $\sqrt{2}$ as the photometric precision, which is demonstrated in Fig. \ref{fig:magerr}.
We also obtained the limiting magnitude at $\sigma=0.2$. 
For bright stars, the photometric precision with a 6 pixels aperture reaches 8 mmag in a single frame, which could be further improved by stacking images.

\begin{figure}
    \centering
    \includegraphics[width=\linewidth]{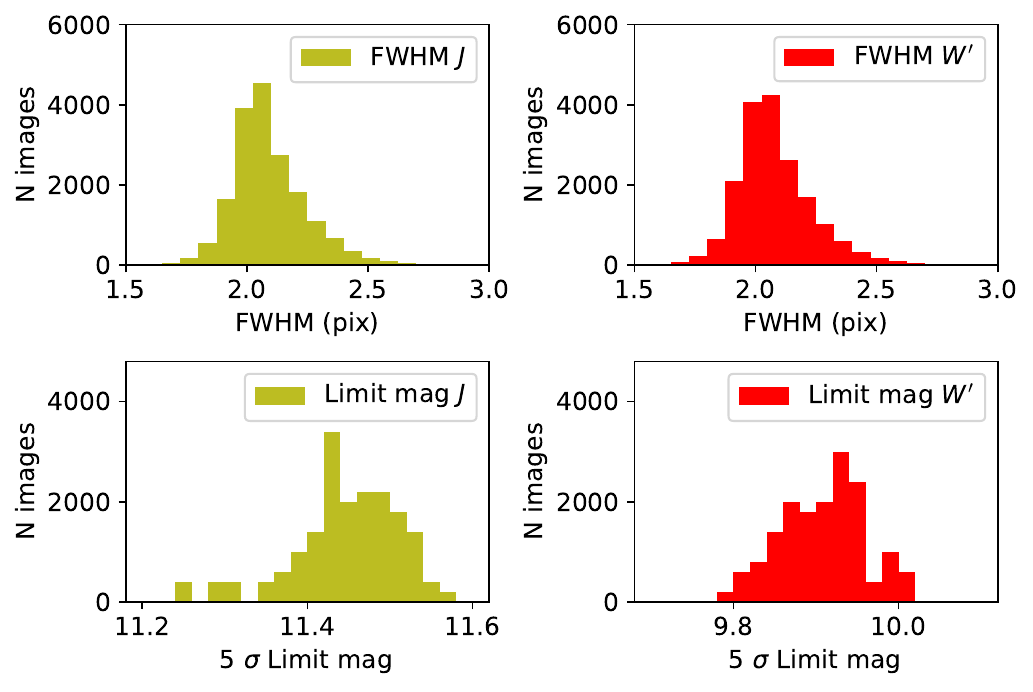}
    \caption{Histograms of the FWHM (upper) and the 5 $\sigma$ limiting magnitudes (lower) for $J$ (yellow) and $W'$ (red) bands, respectively.
            The peaks of FWHM are close to 2 pix.
            The peaks of 5 $\sigma$ limiting magnitudes
            are 11.5 mag for the J band and 9.9 mag for the $W'$ band.}
    \label{fig:hist}
\end{figure}

For all 20,000 images in the EDR, 
we plotted histograms of the FWHM and 5 $\sigma$ limiting magnitudes for $J$ and $W'$ bands in Fig. \ref{fig:hist}.
The median FWHMs are 2.07 and 2.06 pixels for $J$ and $W'$ bands, respectively.
The image depth peaks at $J \thicksim 11.5$ mag and $W' \thicksim 9.9$ mag.
The differences of image quality between the images are marginal, demonstrating that both AIRBT and the site conditions at Dome A are stable.

\subsection{Color-color Diagram}
\label{sec:color}

\begin{figure*}
    \centering
    \includegraphics[width=\linewidth]{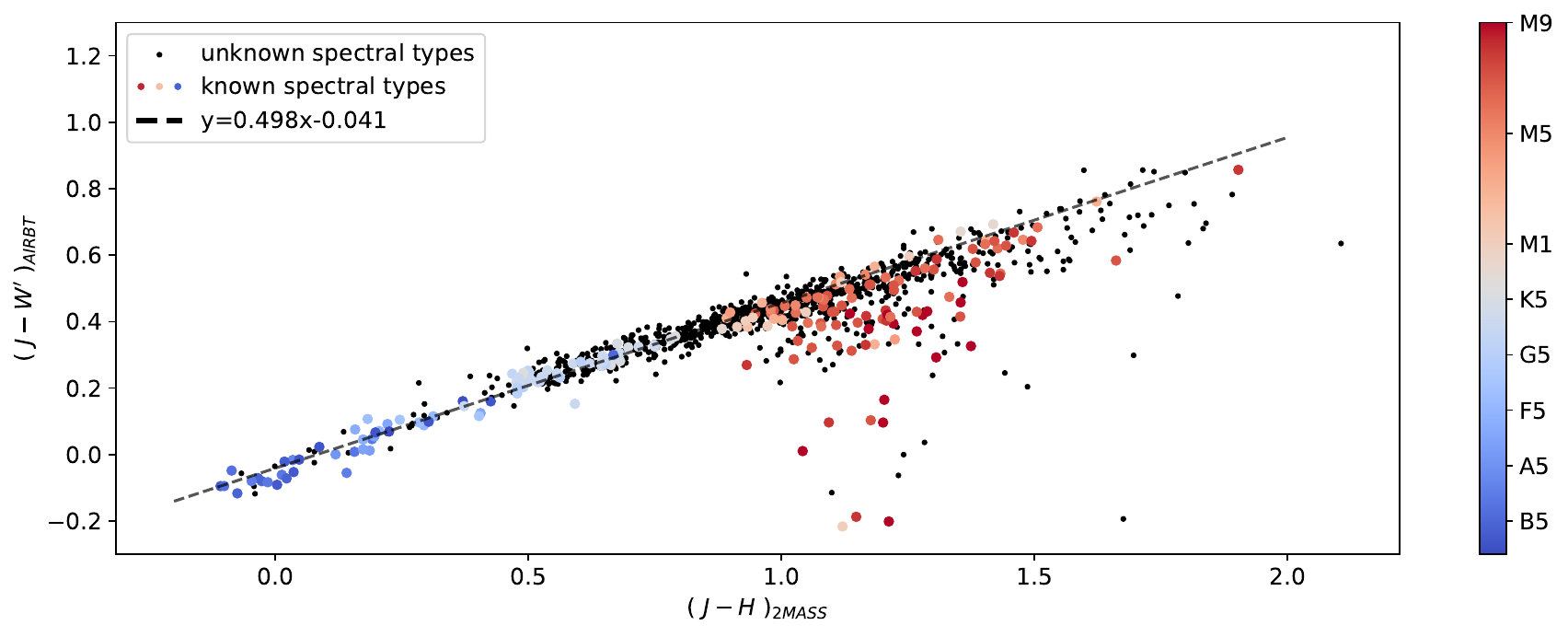}
    \caption{Color-color diagram of $(J-W')_{AIRBT}$ vs $(J-H)_{2MASS}$ from AIRBT data on 3 April 2025.
            The colored points mark stars with known spectral types from SIMBAD ($\thicksim$ 10\%),
            and the black points are stars without spectral type information ($\thicksim$ 90\%).
            The black dashed line is the best fit for AIRBT sources in the range $0<J-H<1$.}
    \label{fig:color}
\end{figure*}

To characterize water-vapor-absorption features near 1.4 \textmu m,  we plot the $(J-W')_{AIRBT}$ vs $(J-H)_{2MASS}$  color-color diagram in Fig. \ref{fig:color}.
Most sources reside along a line, indicating relatively smooth spectra between the $J$ and $H$ bands. The line has a slope of 0.498, slightly steeper than that in Eq. \ref{eq:W_fit} due to the difference of $J$ between AIRBT and 2MASS as described in Sect. \ref{sec:instrument}. However, a small population deviates downward,
implying that their $W'$ magnitudes are larger than expected and thus they are candidates with water-vapor-absorption features.

To verify this hypothesis, we queried the SIMBAD database for the spectral types of all sources and obtained about 10\% with known spectral types \citep{2000A&AS..143....9W, 1994yCat.3170....0M}.
We color-coded the sources according to their spectral types in Fig. \ref{fig:color}.
It is clear that most early type stars (with no absorption features) reside along the line, while most outliers are late M-type stars (probably with absorption features).
The color-color diagram is more robust than $J-H$ color alone, which suffers from varying reddening from interstellar extinction.

It is worth emphasizing that the magnitudes from the two bands for each color should be observed simultaneously, otherwise variability of the sources could possibly introduce significant errors as discussed in the next subsection. Therefore, we adopted $J-W'$ from AIRBT and $J-H$ from 2MASS, but did not use $W'-H$ because there was no $H$ in AIRBT in 2025. It is also the reason we did not use the reddening-independent index $Q$ \citep[e.g.][]{allers2020novel}, which requires simultaneous $J$, $H$, and $W$ magnitudes.

\subsection{Variability in $J$ and $W'$ bands}
\label{sec:JW variation}

\begin{figure}
    \centering
    \includegraphics[width=\linewidth]{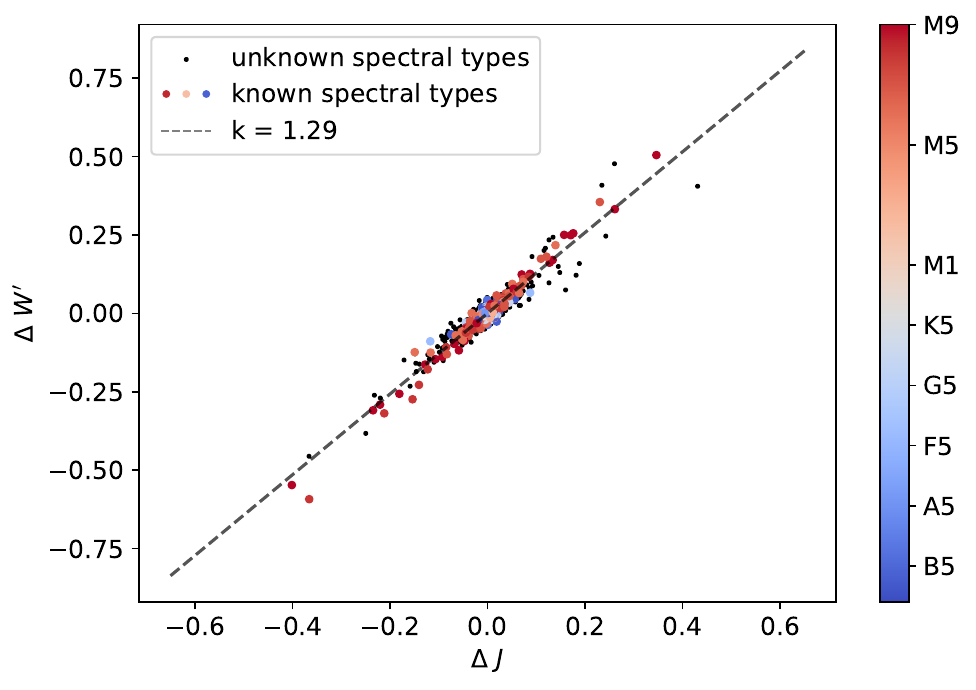}
    \caption{Magnitude variations of $J$ and $W'$ bands between each pair of the three nights of all stars.
            The colored points mark known stars from SIMBAD ($\thicksim$ 10\%),
            and the black points are stars without spectral type information ($\thicksim$ 90\%).
            The black dotted line is the best fit of stars in the range $\left\lvert \Delta J \right\rvert >0.05$ and $\left\lvert \Delta W' \right\rvert >0.05$ ($\thicksim 5\%$ of the stars).}
    \label{fig:mag_compare}
\end{figure}

We compared $\Delta J$ and $\Delta W'$ between each pair of the three nights
(16 March, 3 April, and 20 April 2025) for all sources, 5\% of which had significant variations ($>$ 0.05 mag).
Fig.~\ref{fig:mag_compare} evidently shows that most variable stars exhibit larger variations in $W'$ band and they are distributed around:
\begin{equation}
\begin{aligned}
   \Delta W' \thicksim 1.29 \times \Delta J.
\end{aligned}
\label{f:W_vs_J_obs}
\end{equation}
Furthermore, some known M-type stars tend to show even steeper slopes.

However, variability normally decreases with increasing wavelength, as can be understood from the following argument.
For continuum flux, variations are mainly driven by radius and temperature.
According to the Stefan-Boltzmann law, radius variations affect $J$ and $W'$ bands equally, satisfying $\Delta J = \Delta W'$.
On the other hand, the effect of temperature variation is given by Planck's law, which yields $\Delta W' \thicksim 0.88 \Delta J$ for 1000 K $< T <$ 4000 K.
Therefore, when both radius and temperature variations are present, we should have $\Delta W' < \Delta J$.

\begin{figure}
    \centering
    \includegraphics[width=\linewidth]{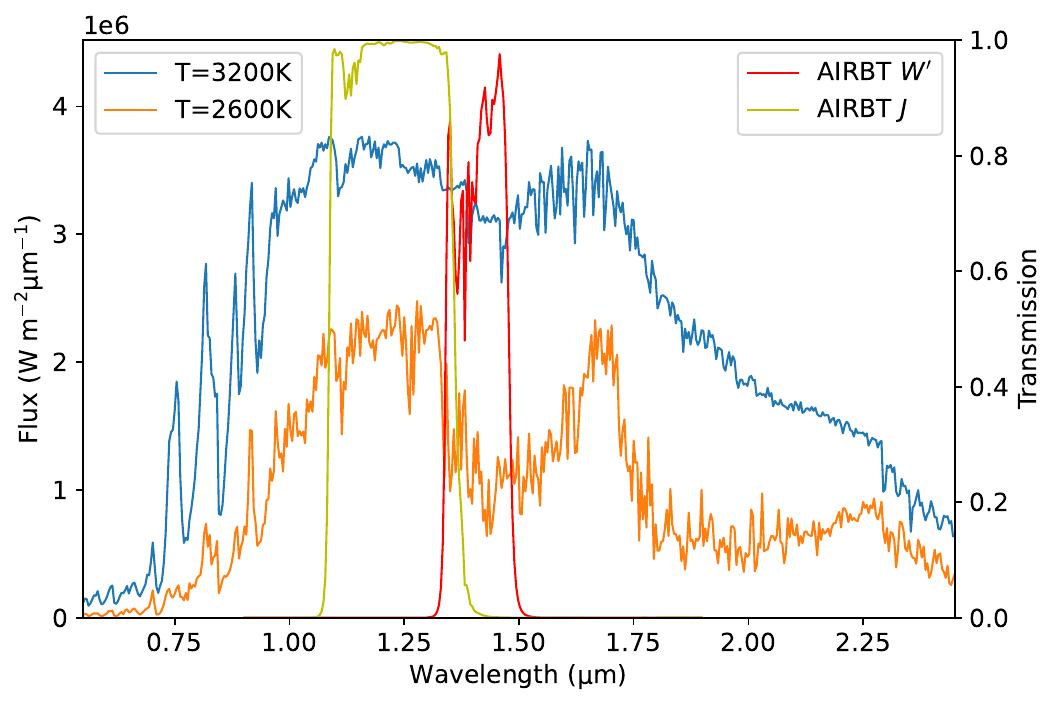}
    \caption{Spectrum variability from the PHOENIX model with $\log{g}=1$ and $[M/H]=0.0$.
            The blue (orange) line shows spectra of a giant star with temperatures of 3200K (2600K),
            and the red (yellow) line is the transmission curve of the AIRBT $W'$ ($J$) filter.
            The $W'$ filter covers the water-vapor-absorption feature near 1.4 \textmu m.}
    \label{fig:Phoenix}
\end{figure}

We propose that the excess $\Delta W'$ comes from variations of water-vapor in the star's atmosphere, thereby
causing the absorption rates near 1.4 \textmu m to vary.
To test this, we used synthetic spectra from the PHOENIX model
\citep{husser2013new}
to simulate the relationship between $\Delta J$ and $\Delta W'$ produced by temperature variations.
Since most of our samples are giants from the Milky Way, we adopted the model parameters for them: $\log{g}=1$ and $[M/H]=0.0$.
As shown in Fig. \ref{fig:Phoenix}, the strength of water-vapor-absorption feature near 1.4 \textmu m does vary with temperature.
The simulations yield $\Delta W'= (1.4 \thicksim 2.5) \Delta J$ for M-type giants with 2500 K $< T <$ 3200 K, which is consistent with our results.
Since our observed variable sources include both early and late type stars,
the variation ratio in Eq. \ref{f:W_vs_J_obs} lies between the continuum prediction and the PHOENIX model.
If the ratio of $\Delta W'/\Delta J$ is larger than 1, it signifies a late-type star.

Discrepancies between $\Delta J$ and $\Delta W'$ may also arise from additional effects.
Light curves of ultracool stars would show different phases between $J$ and $W'$ bands
because the formation and dissociation of atmospheric water-vapor lag behind the temperature change.
Additionally, seasonal variations in $W'$ band could only be present from some ultracool stars under specific conditions. The observed $W'$ flux is a convolution of the stellar spectrum and the telluric atmosphere transmission, both of which contain water-vapor-absorption features. The water-vapor produces many deep (close to 100\%) and narrow ($<$ 1 nm) absorption lines, which are already smoothed in Fig. \ref{fig:filter}. The convolution reaches maximum when both spectra line-up perfectly and drops when they are misaligned.
As the apparent radial velocity of the target star varies with Earth's orbital motion, the $W'$ flux fluctuates accordingly. 
We will further investigate these effects with well-sampled light curves in the future work.

\section{Discussion}

In this section, we discuss the scientific applications and potential of the $W'$ band filter.
A more detailed analysis will be presented in a future work.

\label{sec:discussion}

\subsection{Optical depth in the 1.4 \textmu m band}

\begin{figure*}
    \centering
    \includegraphics[width=\linewidth]{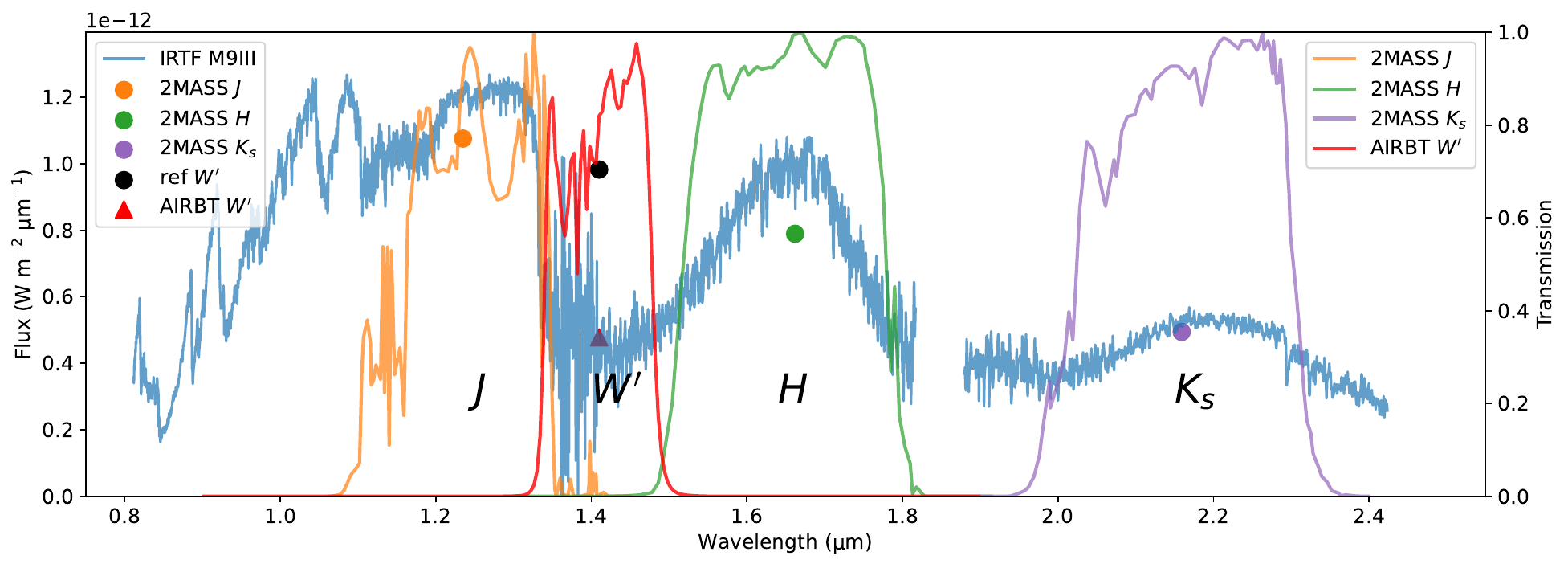}
    \caption{An example showing the calculation of optical depth $\tau$.
    The blue line is a spectrum of an M9 giant BRIB1219-1336 from the IRTF spectral library,
    showing a water-vapor-absorption feature near 1.4 \textmu m.
    The orange, green and purple points are the $J$, $H$ and $K_s$ fluxes converted from 2MASS magnitudes.
    The black point denotes the reference or continuum $W'$ flux from transformation,
    and the red triangle is the $W'$ flux obtained by convolving the spectrum with the filter transmission curve.
    The discrepancy between the expected $W'$ flux and the reference $W'$ flux reflects the water-vapor-absorption feature.}
    \label{fig:m_spec}
\end{figure*}

The strength of water-vapor-absorption can be quantified either by the optical depth $\tau$ or by the reddening-independent index $Q$ \citep[e.g.][]{allers2020novel}.
Here we adopt the optical depth $\tau$ and define it as:
\begin{equation}
    \begin{aligned}
        \tau = \ln{\frac{F_{W', \ cont}}{F_{W'}}} = 0.4 \ln{10} \cdot (W' - W'_{cont}),
    \end{aligned}
    \label{f:tau}
\end{equation}
where $F_{W'}$ and $F_{W',\ cont}$ are the observed $W'$ flux and the continuum $W'$ flux,
as shown in Fig. \ref{fig:m_spec}.

The $W'_{cont}$ is transformed from $J$ and $H$ magnitudes via Eq. \ref{eq:W_fit}.
Since AIRBT lacks its own $H$ band observations, we used the 2MASS $J$ and $H$ magnitudes to conduct the transformation.
However, as the 2MASS and AIRBT observations are separated by a long period of time, we need to correct the variation of $W'_{cont}$, which could be as large as 0.5 mag shown in Fig. \ref{fig:mag_compare}.
As established in Sect. \ref{sec:JW variation}, $\Delta W'_{cont} \leqq \Delta J_{cont}$.
Thus, for the continuum flux without water-vapor-absorption, we assume that $\Delta W'_{cont} \approx \Delta J_{cont}$, and correct the variation for $W'_{cont}$ using
\begin{equation}
\begin{aligned}
    W'_{cont} = W'_{ref} - \Delta W'_{cont} \approx W'_{ref} - (J_{AIRBT} - J_{2MASS}),
\end{aligned}
\label{f:W_cont}
\end{equation}
where $W'_{ref}$ is the reference $W'$ magnitude transformed from 2MASS $J$ and $H$ (Eq. \ref{eq:W_fit}).
For bright stars ($6.5 < J < 9$ and $5.5 < W' < 8$), the uncertainty of $\tau$ is $\thicksim$ 0.05 calculated using error propagation from magnitude errors.

\begin{figure*}
    \centering
    \includegraphics[width=\linewidth]{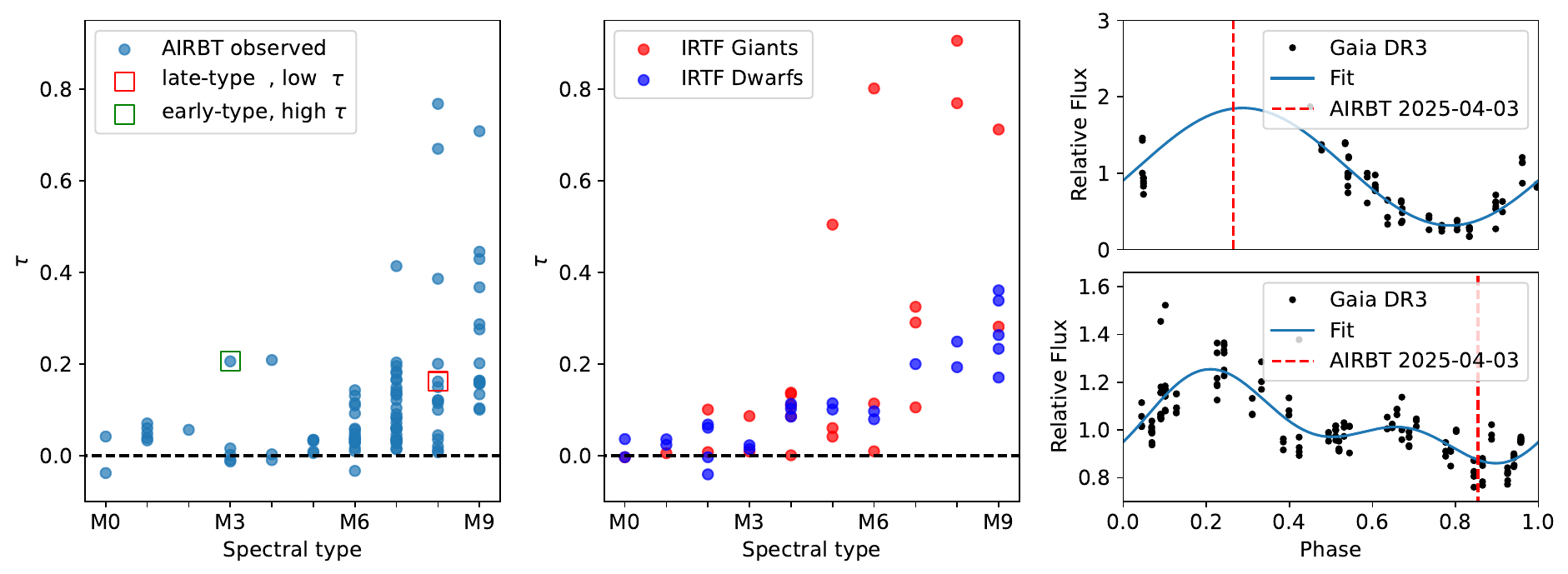}
    \caption{Left: Relationship between optical depth $\tau$ and spectral type from AIRBT known M-type stars.
              Square points mark 2 examples examined in right panel.
              Middle: Results from IRTF spectral library separated into giants (red) and dwarfs (blue).
              Right Upper (Lower) panel: Light curve of the M8 giant Gaia DR3 5255960181168967808 (M3 giant Gaia DR3 5335159343746430336), 
              which was identified as late (early) spectral type but shows a low (high) optical depth,
              and AIRBT observed it at the brighter (fainter) phase indicated by the red dashed line.}
    \label{fig:tau_Gaia}
\end{figure*}

Depressions of the spectrum of ultracool stars near 1.4 \textmu m caused by water-vapor-absorption should be detectable as high optical depth.
In the left panel of Fig. \ref{fig:tau_Gaia}, we plot the optical depth $\tau$ against the spectral type
for the known M-type stars we observed.
For comparison, we also plot $\tau$ calculated from spectra from the IRTF spectral library in the right panel.
Our observational data show a trend similar to that of IRTF giants:
the optical depth $\tau$ is $\approx 0$ for spectral type M0-M5
and increases for later spectral types (M6 and later).

However, there is a large dispersion in the left and middle panels of Fig. \ref{fig:tau_Gaia},
which may be explained by the variation of spectral types, since
most of the sources are Mira variables. 
When a Mira's brightness decreases, its temperature decreases, the spectral type becomes later, and the water-vapor in the atmosphere increases
\citep{boyd2021spectroscopic,smith2002infrared,matsuura2002time}, and vice versa.
Consequently, their spectral types at the epoch of our observation may differ from
those when they were classified as late-type stars by \cite{macconnell2010vizier}.

We examined 2 examples whose optical depth and spectral type (when they were verified) are inconsistent,
as square markers show in the left panel of Fig. \ref{fig:tau_Gaia}:
one was identified as M8 but had a low optical depth,
and the other was identified as M3 but had a high optical depth.
We retrieved their optical light curves from Gaia DR3 \citep{2016A&A...595A...1G}.
Given that the Gaia DR3 light curve spanned 2015--2017 while our observations were conducted in 2025,
we used the Generalized Lomb-Scargle (GLS) method to compute periods
and inferred phases at the epoch of AIRBT observations in the right panel of Fig. \ref{fig:tau_Gaia}.
We find that the inconsistent $\tau$ can be explained by the variability: the M8 one was observed by AIRBT at almost peak brightness, indicating a higher temperature, lower water-vapor-absorption and smaller $\tau$. The M3 one was observed close to the minimum brightness, which also leads to the observed differences in $\tau$.

Thus we can use the optical depth $\tau$ to estimate stellar spectral types, similarly to the method used by \cite{allers2020novel} to estimate spectral types of late-type dwarfs.
However, the fine relationship between $\tau$ and spectral type is affected by several factors,
such as 
luminosity class (supergiants, giants, dwarfs) and stellar parameters.
We will further explore this relationship using an expanded dataset in a future work.

\subsection{Dome A Atmosphere Transmission Stability}

The atmosphere transmission would vary as a result of the fluctuation of PWV.
For PWV values of 0.05 mm (10\%) and 0.268 (90\%) at Dome A by \cite{kuo2017assessments}, the average transmission of $W'$ band is 89.6\% and 75.5\%, respectively \citep{zhang2024customizing}.
They correspond to a magnitude difference of 0.08 mag compared to the median PWV.

\begin{figure}
    \centering
    \includegraphics[width=\linewidth]{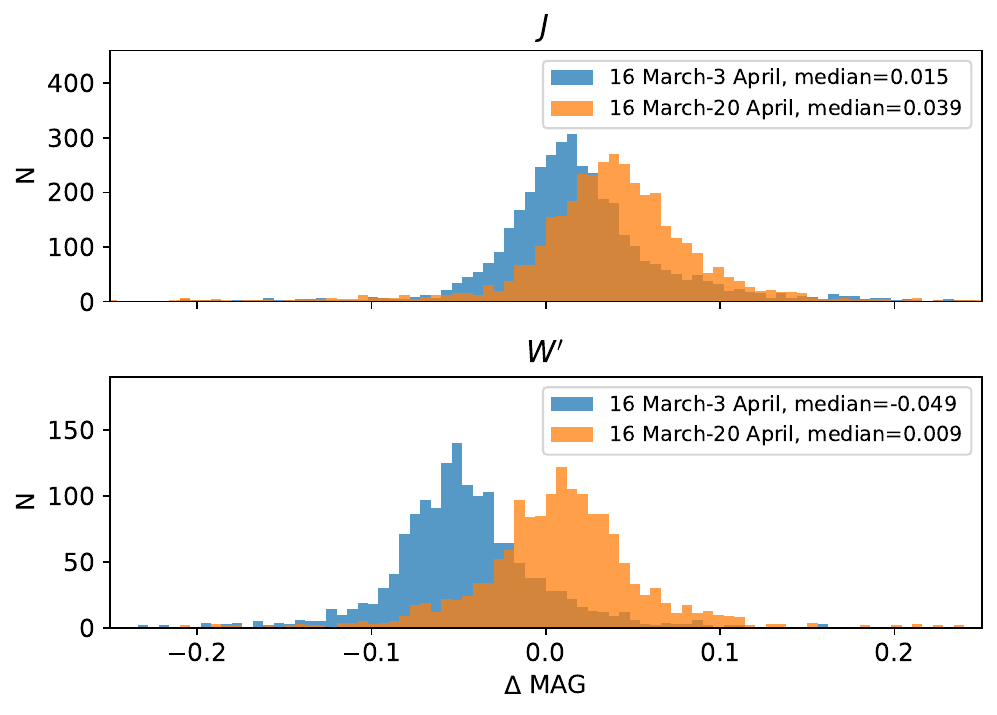}
    \caption{Atmosphere transmission stability for $J$ (upper) and $W'$ (lower) bands at Dome A, Antarctica.
            The blue (orange) histograms are instrumental magnitude differences with a 6 pixels aperture between 16 March and 3 April (20 April) 2025.}
    \label{fig:DomeA-air}
\end{figure}

To verify the atmosphere transmission stability of Dome A, we compared the instrumental magnitudes of stars between 16 March, 3 April, and 20 April 2025.
We selected stars with $SNR > 50$ on each of the three nights, cross-matched them,
and subtracted instrumental magnitudes to obtain differential magnitudes;
the resulting histogram is shown in Fig.~\ref{fig:DomeA-air}.
The median instrumental magnitude difference with a 6 pixels aperture was $<$ 0.05 mag for both $J$ and $W'$ bands, possibly
owing to the variations of atmosphere transmission as well as accumulation of marginal frost on the windows.
This value agrees well with theoretical predictions above.
It demonstrates that the atmosphere transmission for $W'$ band at Dome A is not only high but also stable.

\section{Conclusion}
\label{sec:summary}

To take advantage of high atmospheric transmission at Dome A,
we designed a new $W'$ filter covers the entire water-vapor-absorption region around 1.4 \textmu m and installed it in AIRBT at Dome A in 2025.
We selected simultaneous $J$ and $W'$ images from three nights between March and April 2025 as an Early Data Release (EDR) to test the performance of $W'$ band filter.
We derived a preliminary calibration for the $W'$ band by transforming $J$ and $H$ magnitudes.
And we achieved 5 $\sigma$ limiting magnitudes of $J \thicksim 11.5$ and $W' \thicksim 9.9$ for 2 s exposures.

With $W'$ band, we successfully detected candidates with a water-vapor-absorption feature near 1.4 \textmu m by two methods.
In the $J-W'$ vs $J-H$ color-color diagram, we identified outliers that exhibit lower $W'$ fluxes than expected.
In the variability analysis, most of the variable sources are more variable in $W'$ than in $J$, and we attribute the excess $\Delta W'$ to the variations of the water-vapor-absorption feature.
Furthermore, we quantified the absorption by optical depth, $\tau$, and found $\tau$ increasing with spectral types for M6 and later.
This indicates that our observations can be used to search for ultracool stars and estimate their spectral types.

We verified the atmosphere transmission stability for $W'$ band at Dome A is better than 5\%.
The high stability indicates that Dome A is a promising site for future observations to study water-vapor in the universe.

We will process all $J$ and $W'$ data obtained in 2025 with fine magnitude calibrations.
While most of EDR targets are giants in the Milky Way limited by the single frame depth,
we plan to co-add multiple exposures to increase our detection depth by about 3 mag,
which enables the future detection of fainter ultracool dwarfs.

In the future, we will replace the telescope's equatorial mount to enable stable pointing and tracking. This will allow us to survey a larger area and observe particular targets, including both scientific sources and standard stars.
For the next step, we plan to install a 40 cm telescope equipped with a larger format and lower noise camera.
We will also carry out time-domain observations in $W'$ band at Dome A to study the formation and evolution of water-vapor in the universe.

\begin{acknowledgements}
We are grateful to the 39th, 40th and 41st CHINARE teams supported by the Polar Research Institute of China and the Chinese Arctic and Antarctic Administration.
B.M. acknowledges the support from the National Key R\&D Program of China (grant No. 2022YFC2807303).
We also acknowledge the supports by National Astronomical Observatories, Chinese Academy of Sciences under grant numbers E355350101 and E4TG350101 and by the National Natural Science Foundation
of China under grant numbers 11733007 and 12373092.
AIRBT is financially supported by School of Physics and Astronomy, Sun Yat-sen University.
B.M. acknowledges the support by National Astronomical Data Center, the Greater Bay Area No. 2024B1212080003.
\end{acknowledgements}

\software{astropy \citep{2013A&A...558A..33A,2018AJ....156..123A,2022ApJ...935..167A}, SEP \citep{Barbary2016}, SExtractor \citep{1996A&AS..117..393B}, SCAMP\citep{2006ASPC..351..112B},TOPCAT \citep{2005ASPC..347...29T}}

\begin{appendix}

\section*{Catalog}
The final EDR covers $\alpha = 145 \thicksim 205^\circ$ and $\delta = -61.7 \thicksim -61.0^\circ$ in both $J$ and $W'$ bands.
It consists of three catalogs, one for each night on 16 March, 3 April, and 20 April 2025.
Each catalog combines photometric results from $\thicksim$ 6800 images in $J$ and $W'$ bands.
For every source within a night, we take the median photometric results and record the total number of detections in individual frames.
Each night includes $\thicksim$ 15000 identical sources that have more than 30 measurements, corresponding to effective exposure times longer than one minute, to ensure data validity.

The detailed parameters of the catalog are listed in Table \ref{tab:catalog}.
The catalog can be accessed via \url{https://nadc.china-vo.org/res/r101738/}.

\begin{table*}[h!]
\caption{Content of the objects in catalog}                 
\label{tab:catalog}    
\centering 
\renewcommand\arraystretch{1.0}
\begin{tabular}{| l | c | c | l |}
\hline               
Column Name & Type & Unit & Description \\
\hline                      
    RAJ2000       & float64           & deg           & median RA (J2000) of all the source detections in this night \\
    \hline
    DEJ2000       & float64           & deg           & median Dec (J2000) of all the source detections in this night \\
    \hline

    Jmag\_APER    & float32[3]        & mag           & array of $J$ band photometry MAG\_APER at aperture 3,6,8 pixel \\
    \hline
    e\_Jmag\_APER & float32[3]        & mag           & array of the uncertainty of $J$ band photometry MAG\_APER at aperture 3,6,8 pixel \\
    \hline
    FWHM\_J       & float32           & pixel         & $J$ band median fwhm of all the source detections in this night \\
    \hline
    MJD\_J        & float64           & -             & $J$ band median MJD (UTC) of all the source detections in this night \\
    \hline
    AWIN\_J       & float32           & pixel           & $J$ band median long axis of all the source detections in this night \\
    \hline
    BWIN\_J       & float32           & pixel           & $J$ band median short axis of all the source detections in this night \\
    \hline
    errAWIN\_J    & float32           & pixel           & uncertainty of $J$ band median long axis of all the source detections in this night \\
    \hline
    errBWIN\_J    & float32           & pixel           & uncertainty of $J$ band median short axis of all the source detections in this night \\
    \hline
    N\_J          & int16             & -             & detections of this source in this night of $J$ band \\
    \hline
    %
    Wmag\_APER    & float32[3]        & mag           & array of $W'$ band photometry MAG\_APER at aperture 3,6,8 pixel \\
    \hline
    e\_Wmag\_APER & float32[3]        & mag           & array of the uncertainty of $W'$ band photometry MAG\_APER at aperture 3,6,8 pixel \\
    \hline
    FWHM\_W       & float32           & pixel         & $W'$ band median fwhm of all the source detections in this night \\
    \hline
    MJD\_W        & float64           & -             & $W'$ band median MJD (UTC) of all the source detections in this night \\
    \hline
    AWIN\_W       & float32           & pixel           & $W'$ band median long axis of all the source detections in this night \\
    \hline
    BWIN\_W       & float32           & pixel           & $W'$ band median short axis of all the source detections in this night \\
    \hline
    errAWIN\_W    & float32           & pixel           & uncertainty of $W'$ band median long axis of all the source detections in this night \\
    \hline
    errBWIN\_W    & float32           & pixel           & uncertainty of $W'$ band median short axis of all the source detections in this night \\
    \hline
    N\_W          & int16             & -             & detections of this source in this night of $W"$ band \\
    \hline
    RAJ2000\_2MASS     & float32      & deg           & RA (J2000) of the brightest 2MASS source within 3 pixel \\
    \hline
    DEJ2000\_2MASS     & float32      & deg           & DEC (J2000) of the brightest 2MASS source within 3 pixel \\
    \hline
    Jmag\_2MASS   & float32           & mag           & 2MASS $J$ band magnitude compute by summarizing flux within 3 pixel \\
    \hline
    Hmag\_2MASS   & float32           & mag           & 2MASS $H$ band magnitude compute by summarizing flux within 3 pixel \\
    \hline
    Kmag\_2MASS   & float32           & mag           & 2MASS $K$ band magnitude compute by summarizing flux within 3 pixel \\
    \hline
    Wmag\_REF     & float32           & mag           & reference $W'$ magnitude transformed by 2MASS magnitude \\
    \hline
    e\_Jmag\_2MASS  & float32         & mag           & uncertainty of 2MASS $J$ band magnitude compute by summarizing flux within 3 pixel \\
    \hline
    e\_Hmag\_2MASS  & float32         & mag           & uncertainty of 2MASS $H$ band magnitude compute by summarizing flux within 3 pixel \\
    \hline
    e\_Kmag\_2MASS  & float32         & mag           & uncertainty of 2MASS $K$ band magnitude compute by summarizing flux within 3 pixel \\
    \hline
    e\_Wmag\_REF    & float32         & mag           & uncertainty of reference  $W'$ magnitude transformed by 2MASS magnitude \\
    \hline

\end{tabular}
\end{table*}

\end{appendix}

\bibliography{sample7}{}
\bibliographystyle{aasjournalv7}

\end{CJK*}
\end{document}